# Valley edge states as bound states in the continuum

Shunda Yin[1†], Liping Ye[1†], Hailong He[1], Xueqin Huang[2], Manzhu Ke[1], Weiyin Deng[1*], Jiuyang Lu[1*], Zhengyou Liu[1,3*]

[1]Key Laboratory of Artificial Micro- and Nanostructures of Ministry of Education and School of Physics and Technology, Wuhan University, Wuhan 430072, China

[2]School of Physics and Optoelectronics, South China University of Technology, Guangzhou, 510640, China

[3]Institute for Advanced Studies, Wuhan University, Wuhan 430072, China

†These authors contributed equally to this work.
*Corresponding author. Emails: dengwy@whu.edu.cn; jylu@whu.edu.cn; zyliu@whu.edu.cn

Bound states in the continuum (BICs) are spatially localized states with energy embedded in the continuum spectrum of extended states. The combination of BICs physics and nontrivial band topology theory giving rise to topological BICs, which are robust against disorders and meanwhile of the merit of conventional BICs, is attracting wide attention recently. Here, we report valley edge states as topological BICs, which appear at domain wall between two distinct valley topological phases. The robustness of such BICs is demonstrated. The simulations and experiments show great agreement. Our findings of valley related topological BICs shed light on both BICs and valley physics, and may foster innovative applications of topological acoustic devices.

Bound states in the continuum (BICs) are resonances with zero leakage featured by discrete energies embedded in a continuum background spectrum [1-3]. BICs have been realized via various mechanisms, such as symmetry incompatibility [4-6] and parameter tuning [7-10], in different systems [9-18]. Among them, phononic crystal (PC) for acoustic wave acts as a powerful platform to explore the BICs, even in the presence of open non-Hermitian interactions, because of its extraordinary controllability and tunability [19,20]. Benefited from the high-$Q$ values and strong localizations, BICs have been proposed for extensive applications, including large-area high-power lasers in photonic slabs [21,22], chemical and biological sensing [23], and supersonic surface acoustic wave filters [24].

In recent years, topological matter has emerged as a booming and fruitful branch [25-28]. The fundamental feature of topological systems is the existence of the topological boundary states predicted by bulk-boundary correspondence. Very recently, the combination of BICs and nontrivial band topology leads to a new type of states, referred to as topological BICs [29-37]. Specifically, topological BICs are the topological boundary states with their energies survived in the continuum spectrum, where the hybridization between topological bound states and background extended states is forbidden. In the presence of topological protection, such BICs are robust against imperfections and disorders. Topological BICs are extensively studied in higher-order topological systems [38,39], such as the zero-dimensional (0D) corner states in 2D systems [31-36] and the 1D hinge states in 3D systems [37]. However, these states are nondispersive and thus can hardly propagate. For dispersive BICs, the topological boundary states are propagatable in a continuum and keep highly confined in transverse directions, beyond the conventional case where the topological states only exist in bandgaps.

In this work, we realize valley topological BICs (VTBICs), which inherit dispersive propagations from valley topology and keep highly confined due to the mechanism of separability. VTBICs are constructed in 2D systems by coupling two identical layers of valley topological insulators [40-46]. The associated Hamiltonian can be separated into two orthogonal topological subsystems. In each subsystem, the spatially confined topological states occur at the interface between two topological phases with opposite valley Chern numbers. These valley edge states form BICs since the edge states of one subsystem appear in bulk continuum of the other subsystem without hybridization, and vice versa. Note that the proposed VTBICs here are dispersive and can propagate robustly along the interfaces in the background spectrum, significantly different from the previous reported nondispersive topological BICs [31-36]. VTBICs and their reflection immunities have been conclusively identified by performing airborne sound experiments in PCs.

We start by considering two identical layers of valley topological insulators constructed in a square lattice and coupled indirectly through a middle layer (with the sites colored in blue), as shown in Fig. 1(a). For this sandwiched structure, each unit cell includes five sites labeled by numbers, and the lattice constant of the square lattice is $a/\sqrt{2}$ (with $a=1$ for simplicity). The on-site energies for the green and pink sites are $m$ and $-m$, respectively, and for the blue ones are $m_5$. The hopping strengths for intralayer couplings are labeled $t_a$ and $t_b$, and for interlayer ones are $t_1$ and $t_2$. The Hamiltonian of this model is thus written as

$$H_k = \begin{pmatrix} m & w & 0 & 0 & 0 \\ w^* & -m & 0 & 0 & t_1 \\ 0 & 0 & m & w & 0 \\ 0 & 0 & w^* & -m & t_2 \\ 0 & t_1 & 0 & t_2 & m_5 \end{pmatrix}, \quad (1)$$

where $w = t_a + 2t_b \cos(k_x/2) e^{-ik_y/2} + t_b e^{-ik_y}$. It shows that without the interlayer couplings ($t_1$ and $t_2$) the Hamiltonian $H_k$ is separated in three independent layers. In fact, even in the presence the interlayer couplings, the Hamiltonian $H_k$ can be separated into two independent subsystems: after a unitary transformation with

$$U = \begin{pmatrix} -\cos\phi & 0 & \sin\phi & 0 & 0 \\ 0 & -\cos\phi & 0 & \sin\phi & 0 \\ \sin\phi & 0 & \cos\phi & 0 & 0 \\ 0 & \sin\phi & 0 & \cos\phi & 0 \\ 0 & 0 & 0 & 0 & 1 \end{pmatrix}, \quad (2)$$

$H_k$ is block-diagonalized as $H'_k = U^\dagger H_k U = \begin{pmatrix} h^{(2)} & 0 \\ 0 & h^{(3)} \end{pmatrix}$, where $\tan\phi = t_2/t_1$,

$$h^{(2)} = \begin{pmatrix} m & w \\ w^* & -m \end{pmatrix}, \text{ and } h^{(3)} = \begin{pmatrix} m & w & 0 \\ w^* & -m & \sqrt{t_1^2 + t_2^2} \\ 0 & \sqrt{t_1^2 + t_2^2} & m_5 \end{pmatrix}. \tag{3}$$

Here, $h^{(2)}$ and $h^{(3)}$ are the Hamiltonians of two subsystems in subspaces of dimensions 2 and 3, respectively. The $h^{(2)}$ subsystem involves only the upper and lower layers, while the $h^{(3)}$ subsystem further is relevant to all the three layers. Note that, the block diagonalization procedure of $H_k$ holds for arbitrary interlayer couplings $t_1$ and $t_2$.

Figure 1(b) gives the bulk dispersions along the high symmetry lines in the Brillouin zone [right panel of Fig. 1(a)]. Without loss of generality, the interlayer couplings are chosen to be unequal. The dispersions denoted by red and blue curves are derived from the two blocks $h^{(2)}$ and $h^{(3)}$, respectively. Since the unitary transformation generates the upper block $h^{(2)}$ only by mixing the upper and lower layers, the wave functions associated with the red curves disappear on the middle layer (Supplemental S-I [47]). Actually, the upper block $h^{(2)}$ has identical form with that of the single upper or lower layer in the absence of interlayer couplings. Thus, the upper block $h^{(2)}$ possesses the same topology as the valley topological insulator in square lattice [46]. Here we present the nontrivial Berry curvature distribution of the first band of the upper block $h^{(2)}$ in Fig. 1(c), while those of other bands are provided in Supplemental S-II [47]. As shown in Fig. 1(c), Berry curvatures mainly locate around the points $N$ and $N'$ with opposite magnitudes, forming two valleys of time-reversal counterparts. The valley Chern numbers of valleys $N$ and $N'$ can be recognized as $C_N^{(2)} = -1/2$ and $C_{N'}^{(2)} = 1/2$. Significantly different from the case in Ref. [29], the remaining lower block $h^{(3)}$ is also topological nontrivial with the valley Chern number being $(C_N^{(3)}, C_{N'}^{(3)}) = (1/2, -1/2)$.

A topologically distinct phase can be constructed by changing the on-site energies of five sites and forming a new unit cell shown in Fig. 1(d). Compared to the phase in Fig. 1(a), the on-site energies of sites 1 and 3 are interchanged with those of sites 2 and 4, and sites 1 and 3 are coupled indirectly via site 5. For this new phase, the valley Chern numbers of both subsystems are inversed with respect to those of the phase in Fig. 1(a), while the bulk dispersions remain the same (see Supplemental S-I and S-II [47]). Hereafter, we denote the topological phases in Figs. 1(a) and 1(d) as phases I and II, respectively. To investigate the valley edge states, these two phases are placed together to form a supercell as illustrated in Fig. 1(e). The bulk-boundary correspondence guarantees the existence of spatially confined valley topological states occurring at the interface and each subsystem of phases I and II with inversed valley Chern numbers host its own valley edge states. Specifically, as shown in Fig. 1(f), the $h^{(2)}$ subsystem (with the bulk states denoted by red shadows) induces the valley edge states denoted by magenta curve, and the $h^{(3)}$ subsystem (with the bulk states denoted by blue shadows) hosts the valley edge states denoted by cyan curve. More importantly, the valley edge states of the $h^{(2)}$ subsystem can survive in the bulk states of the $h^{(3)}$ subsystem and vice versa. The hybridization of bulk and valley edge states is forbidden due to the subsystems are orthogonal. This is further verified by the eigenstate distributions of valley edge states shown in the right panel of Fig. 1(f), where the valley edge states are highly confined at the interface and do not extend into the bulk. Thus, the valley edge states here are the VTBICs. As the VTBICs inherit from the bulk states, VTBICs originated from the $h^{(2)}$ subsystem vanish on the middle layer even in the absence of mirror symmetry about the middle layer, while those of the $h^{(3)}$ subsystem are distributed at all layers. These properties facilitate our experimental observations below. Moreover, one can tune the parameters to obtain different types of VTBICs, as discussed in Supplemental S-III [47].

We now show that the VTBICs elaborated above can be realized in PCs. The tight-binding model of Fig. 1(e) can be emulated by an acoustic cavity-tube structure directly, as shown in Fig. 2(a). The unit cell (side and top views) of phase I is shown in Fig. 2(b). The five acoustic cavities (tetragonal prism) mimic the sites, and the cylinders emulate the couplings among them. Specifically, the intralayer couplings are emulated by the gray and black cylinders with diameters $d_1 = 2$ mm, and $d_2 = 8$ mm, respectively, and the interlayer couplings are emulated by the yellow cylinders with radius $r_0 = 4.8$ mm. Here, we choose the equal interlayer couplings for simplicity. In this case, the wave functions in the $h^{(2)}$ and $h^{(3)}$ subsystems are antisymmetric and symmetric (Supplemental S-IV [47]), respectively, which facilitate our experimental observations. Note that the states in the $h^{(2)}$ and $h^{(3)}$ subsystems have no parity symmetry in the general case. The other geometrical parameters are chosen as $a = 24$ mm, $s =$ 10 mm, $l_0 = 1.8$ mm, $l_1 = 11$ mm, $l_2 = 16$ mm, $l_5 = 10$ mm. Similarly, the unit cell of phase II can be constructed by changing the heights of five cavities and forming a new unit cell. We have checked that bulk properties (bulk dispersions, symmetries and topology of bulk states) of phase I and phase II are consistent well with the predictions, where $h^{(2)}$ and $h^{(3)}$ subsystems have band gaps in the range of 7.47-8.14 kHz and 8.17-8.78 kHz, respectively (Supplemental S-IV [47]). The projected dispersions for the interface structure in Fig. 2(a) are presented in Fig. 2(c). As expected, the acoustic VTBICs (solid magenta curve) originated from $h^{(2)}$ subsystem appear in the continuum spectrum of the extended bulk states (blue dots) of the $h^{(3)}$ subsystem without hybridization, and vice versa. The spatial distributions of these acoustic VTBICs are consistent with the results of tight-binding model, as shown in Supplemental S-V [47].

Based on the fact that the states from the $h^{(2)}$ and $h^{(3)}$ subsystems are not hybridized, the topological states of two subsystems (i.e., bulk states and VTBICs) of the PC can be experimentally validated (Supplemental S-VI [47]). We first identify the bulk states of the $h^{(2)}$ and $h^{(3)}$ subsystems of phase I by the anti-phase and in-phase excitations, respectively. The anti-phase and in-phase sources are realized by a pair of sources with $\pi$ and 0 phase shifts, respectively. By performing 1D Fourier transform of the measured pressure at the positions marked by the red dots in Fig. 2(a), we have mapped out the projected dispersions for the bulk states of $h^{(2)}$ subsystem of phase I. As shown in Fig. 2(c), the experimental data (bright color) agree well with the simulated results (white dots), where the band gap of $h^{(2)}$ subsystem is clearly presented. Similarly, the in-phase sources can well excite the bulk states of $h^{(3)}$ subsystem and the measured projected dispersions (bright color) agrees well with the simulated one (blue dots), as shown in Fig. 2(d). We have checked that the measured projected dispersions of two subsystems of phase II are the same as that of phase I, consistent with the simulations, see Supplemental S-VI [47].

We then turn to the experimental demonstration of acoustic VTBICs. As shown in Fig. 3(a), the measured dispersions exhibits clearly a 1D mode with a positive slope by using anti-phase sources, which reproduces perfectly the simulated dispersions (magenta curve), except for the band broadening due to the finite-size effect. Note that there is no signal of the overlapped bulk states from the $h^{(3)}$ subsystem and negative slope VTBICs from the $h^{(2)}$ subsystem, which further demonstrates the well excitation of the VTBICs from the $h^{(2)}$ subsystem propagating toward $x$ direction. To further characterize the VTBICs from the $h^{(2)}$ subsystem, we present its spatial acoustic pressure distributions in Fig. 3(b), where the sound signals are strong localized at the interface and decay into the bulk. Agreeing with the theoretical prediction, the acoustic fields vanish at the middle layer and are antisymmetric between the upper and lower layers. When using in-phase sources, the measured dispersions and spatial distributions of the VTBICs from the $h^{(3)}$ subsystem are obtained in Figs. 3(c) and 3(d), respectively. Similarly, we have also identified the VTBICs moving along the $-x$ direction by injecting sound waves from the right side of the sample, see Supplemental S-VI [47]. All these experimental data demonstrate the existence of VTBICs of one subsystem embedded in the continuum spectrum of bulk states of the other subsystem.

Below we demonstrate the negligibly weak backscattering of the VTBICs propagating along sharply twisted interfaces. Figure 4(a) shows the sound transport in a Z-shaped bending channel. As exemplified by the field pattern simulated at 7.90 kHz (top) for VTBIC from the $h^{(2)}$ subsystem and 8.55 kHz (bottom) for VTBIC from the $h^{(3)}$ subsystem, respectively, the sound travels smoothly in the curved path despite suffering two sharp corners (bent by 135°). The reflection immunity of VTBICs against sharp corners can be further checked from the corresponding Fourier spectra shown in the insets of Fig. 4(a). The spatial Fourier spectra show that the backward propagating modes are suppressed. The consistence exhibited in the broadband spectra for VTBICs from the $h^{(2)}$ and $h^{(3)}$ subsystems can be seen in Fig. 4(b). All the data confirm the weak influence of the bending corners on the wave transport of the VTBICs.

In conclusion, we have theoretically proposed and experimentally verified the VTBICs in the 2D PC. The acoustic VTBICs, attributed to the valley edge states of two orthogonal subsystems, can be selectively accessed according to their symmetries. We emphasize that the VTBICs can exist for arbitrary interlayer couplings, even in absence of the mirror symmetry of the system in the $z$ direction. Dramatically, we have demonstrated that the acoustic VTBICs can be guided along sharply twisted interfaces with negligible scattering. The VTBICs discussed in our acoustic system can also be realized in electronic, electromagnetic and mechanical systems as well.

**Acknowledgements**

This work is supported by the National Key R&D Program of China (Nos. 2022YFA1404500, 2022YFA1404900), National Natural Science Foundation of China (Nos. 12074128, 12004286, 12104347, 12222405, 12374419, and 12374409), and Guangdong Basic and Applied Basic Research Foundation (Nos. 2021B1515020086, 2022B1515020102).

**References**

[1] J. von Neumann and E. Wigner, Über merkwürdige diskrete Eigenwerte, Phys. Z. **30**, 465 (in German) (1929).
[2] C. W. Hsu, B. Zhen, A. D. Stone, J. D. Joannopoulos, and M. Soljačić, Bound states in the continuum, Nat. Rev. Mater. **1**, 16048 (2016).
[3] M. Kang, T. Liu, C. T. Chan, and M. Xiao, Applications of bound states in the continuum in photonics, Nat. Rev. Phys. **5**, 659 (2023).
[4] E. N. Bulgakov and A. F. Sadreev, Bound states in the continuum in photonic waveguides inspired by defects, Phys. Rev. B **78**, 075105 (2008).
[5] N. Moiseyev, Suppression of feshbach resonance widths in two-dimensional waveguides and quantum dots: A lower bound for the number of bound states in the continuum, Phys. Rev. Lett. **102**, 167404 (2009).
[6] Y. Plotnik, O. Peleg, F. Dreisow, M. Heinrich, S. Nolte, A. Szameit, and M. Segev, Experimental observation of optical bound states in the continuum, Phys. Rev. Lett. **107**, 183901(2011).
[7] S. I. Azzam, V. M. Shalaev, A. Boltasseva, and A. V. Kildishev, Formation of bound states in the continuum in hybrid plasmonic-photonic systems, Phys. Rev. Lett. **121**, 253901 (2018).
[8] A. A. Lyapina, D. N. Maksimov, A. S. Pilipchuk, and A. F. Sadreev, Bound states in the continuum in open acoustic resonators, J. Fluid Mech. **780**, 370 (2015).
[9] S. Huang, T. Liu, Z. Zhou, X. Wang, J. Zhu, and Y. Li, Extreme sound confinement from quasi bound states in the continuum, Phys. Rev. Appl. **14**, 21001 (2020).
[10] L. Huang, Y. K. Chiang, S. Huang, C. Shen, F. Deng, Y. Cheng, B. Jia, Y. Li, D. A. Powell, and A. E. Miroshnichenko, Sound trapping in an open resonator, Nat. Commun. **12**, 4819 (2021).
[11] A. Cerjan, C. Jörg, S. Vaidya, S. Augustine, W. A. Benalcazar, C. W. Hsu, G. von Freymann, M. C. Rechtsman, Observation of bound states in the continuum

embedded in symmetry bandgaps, Sci. Adv. **7**, eabk1117 (2021).

[12] S. Vaidya, W. A. Benalcazar, A. Cerjan, and M. C. Rechtsman, Point-defect-localized bound states in the continuum in photonic crystals and structured fibers, Phys. Rev. Lett. **127**, 023605 (2021).

[13] C. F. Doiron, I. Brener, and A. Cerjan, Realizing symmetry-guaranteed pairs of bound states in the continuum in metasurfaces, Nat. Commun. **13**, 7534 (2022).

[14] M. Kang, Z. Zhang, T. Wu, X. Zhang, Q. Xu, A. Krasnok, J. Han, and A. Alu, Coherent full polarization control based on bound states in the continuum, Nat. Commun. **13**, 4536 (2022).

[15] Y. Chen, H. Deng, X. Sha, W. Chen, R. Wang, Y. Chen, D. Wu, J. Chu, Y. S. Kivshar, S. Xiao, and C. Qiu, Observation of intrinsic chiral bound states in the continuum, Nature **613**, 474 (2023).

[16] C. W. Hsu, B. Zhen, A. D. Stone, J. D. Joannopoulos and M. Soljačić, Brillouin zone folding driven bound states in the continuum, Nat. Commun. **14**, 2811 (2023).

[17] I. Deriy, I. Toftul, M. Petrov, and A. Bogdanov, Bound states in the continuum in compact acoustic resonators, Phys. Rev. Lett. **128**, 084301 (2022).

[18] M. Amrani et al., Experimental evidence of the existence of bound states in the continuum and fano resonances in solid-liquid layered media, Phys. Rev. Appl. **15**, 054046 (2021).

[19] W. Zhu, W. Deng, Y. Liu, J. Lu, H.-X. Wang, Z.-K. Lin, X. Huang, J.-H. Jiang, and Z. Liu, Topological phononic metamaterials, Rep. Prog. Phys. **86**, 106501 (2023).

[20] L. Huang, S. Huang, C. Shen, S. Yves, A. S. Pilipchuk, X. Ni, S. Kim, Y. K. Chiang, D. A. Powell, J. Zhu, Y. Cheng, Y. Li, A. F. Sadreev, A. Alù, and A. E. Miroshnichenko, Acoustic resonances in non-Hermitian open systems, Nat. Rev. Phys. https://doi.org/10.1038/s42254-023-00659-z (2023).

[21] M. Hideki, Y. Susumu, S. Hirohisa, J. Yue, T. Yoshinori, N. Susumu, GaN photonic-crystal surface-emitting laser at blue-violet wavelengths, Science **319**, 445 (2008).

[22] H. Kazuyoshi, L. Yong, K. Yoshitaka, W. Akiyoshi, S. Takahiro, and N. Susumu, Watt-class high-power, high-beam-quality photonic-crystal lasers, Nat. Photonics **8**, 406 (2014).

[23] X. Fan, and I. M. White, Optofluidic microsystems for chemical and biological analysis, Nat. Photonics **5**, 591 (2011).

[24] G. I. Stegeman, Normal-mode surface waves in the pseudobranch on the (001) plane of gallium arsenide, J. Appl. Phys. **47**, 1712 (1976).

[25] A. Bansil, H. Lin, and T. Das, Colloquium: Topological band theory, Rev. Mod. Phys. **88**, 021004 (2016).

[26] G. Ma, M. Xiao, and C. T. Chan, Topological phases in acoustic and mechanical systems, Nat. Rev. Phys. **1**, 281 (2019).

[27] T. Ozawa *et al.*, Topological photonics, Rev. Mod. Phys. **91**, 015006 (2019).

[28] H. Xue, Y. Yang, and B. Zhang, Topological acoustics, Nat. Rev. Mater. 7, 974 (2022).

[29] Y. Xiao, G. Ma, Z. Zhang, and C. T. Chan, Topological subspace-induced bound state in the continuum, Phys. Rev. Lett. **118**, 166803 (2017).

[30] Z. Li, J. Wu, X. Huang, J. Lu, F. Li, W. Deng, and Z. Liu, Bound state in the

continuum in topological inductor-capacitor circuit, Appl. Phys. Lett. **116**, 263501 (2020).

[31] L. Liu, T. Li, Q. Zhang, M. Xiao, and C. Qiu, A universal mirror-stacking approach for constructing topological bound states in the continuum, Phys. Rev. Lett. **130**, 106301 (2023).

[32] X. Ni, M. Weiner, A. Alu, and A. B. Khanikaev, Observation of higher-order topological acoustic states protected by generalized chiral symmetry, Nat. Mater. **18**, 113 (2019).

[33] W. A. Benalcazar and A. Cerjan, Bound states in the continuum of higher-order topological insulators, Phys. Rev. B **101**, 213901 (2020).

[34] A. Cerjan, M. Jürgensen, W. A. Benalcazar, S. Mukherjee, and M. C. Rechtsman, Observation of a higher-order topological bound state in the continuum, Phys. Rev. Lett. **125**, 213901 (2020).

[35] Y. Wang et al., Quantum superposition demonstrated higher-order topological bound states in the continuum, Light Sci. Appl. **10**, 173 (2021).

[36] Z. Hu, D. Bongiovanni, D. Jukić, E. Jajtić, S. Xia, D. Song, J. Xu, R. Morandotti, H. Buljan, and Z. Chen, Nonlinear control of photonic higher-order topological bound states in the continuum, Light Sci. Appl. **10**, 164 (2021).

[37] Z. Pu, H. He, L. Luo, Q. Ma, L. Ye, M. Ke, and Z. Liu, Acoustic higher-order weyl semimetal with bound hinge states in the continuum, Phys. Rev. Lett. **130**, 116103 (2023).

[38] B. Xie, H. X. Wang, X. Zhang, P. Zhan, J. H. Jiang, M. Lu, and Y. Chen, Higher-order band topology, Nat. Rev. Phys. **3**, 520 (2021).

[39] W. A. Benalcazar, B. A. Bernevig, and T. L. Hughes, Quantized electric multipole insulators, Science **357**, 61 (2017).

[40] J. Lu, C. Qiu, L. Ye, X. Fan, M. Ke, F. Zhang, and Z. Liu, Observation of topological valley transport of sound in sonic crystals, Nat. Phys. **13**, 369 (2017).

[41] M. Wang, L. Ye, J. Christensen, and Z. Liu, Valley physics in non-Hermitian artificial acoustic boron nitride, Phys. Rev. Lett. **120**, 246601 (2018).

[42] M. Yan, J. Lu, F. Li, W. Deng, X. Huang, J. Ma, and Z. Liu, On-chip valley topological materials for elastic wave Manipulation, Nat. Mater. **17**, 993 (2018).

[43] J. Lu, C. Qiu, W. Deng, X. Huang, F. Li, F. Zhang, S. Chen, and Z. Liu, Valley topological phases in bilayer sonic crystals, Phys. Rev. Lett. **120**, 116802 (2018).

[44] F. Gao, H. Xue, Z. Yang, K. Lai, Y. Yu, X. Lin, Y. Chong, G. Shvets, and B. Zhang, Topologically protected refraction of robust kink states in valley photonic crystals, Nat. Phys. **14**, 140(2018).

[45] X. He, E. Liang, J. Yuan, H. Qiu, X. Chen, F. Zhao, and J. Dong, A silicon-on-insulator slab for topological valley transport, Nat. Commun. **10**, 872 (2019).

[46] Z. Zhu, X. Huang, J. Lu, M. Yan, F. Li, W. Deng, and Z. Liu, Negative refraction and partition in acoustic valley materials of a square lattice, Phys. Rev. Appl. **12**, 024007 (2019).

[47] See Supplemental Material for more details.

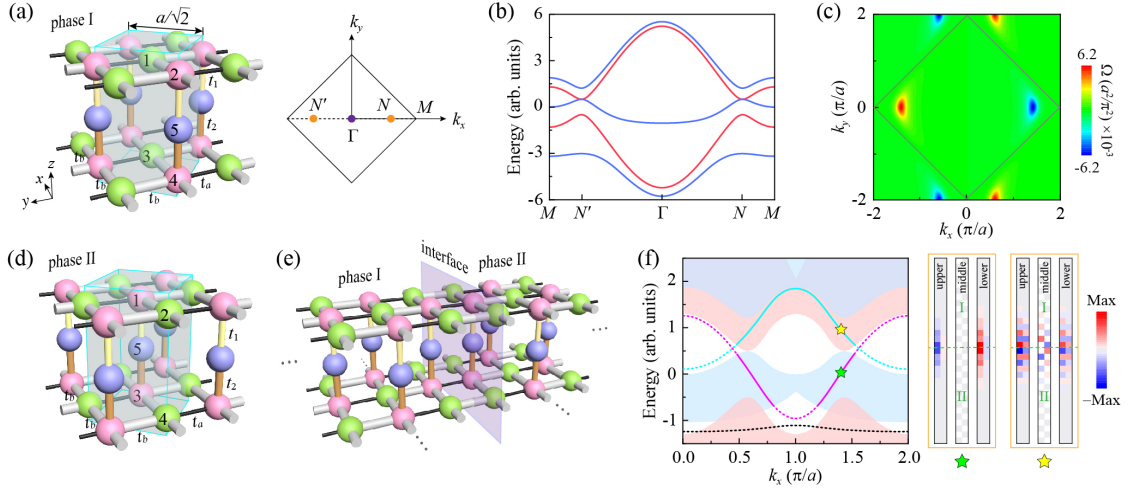

Fig. 1. VTBICs in a tight-binding model with three layers. (a) Left panel: Schematic of the lattice of phase I. Gray box denotes the unit cell with five inequivalent sites labelled 1-5. Right panel: The first Brillouin zone. (b) Bulk dispersions of phase I for $k_y = 0$. The red and blue curves denote the bands of $h^{(2)}$ and $h^{(3)}$, respectively. (c) Berry curvature distribution of the first band of $h^{(2)}$. (d) Lattice of phase II. (e) Interface between phases I and II for realizing the VTBICs. (f) Left panel: Projected dispersions of a ribbon with interface shown in (e). The valley edge states embedded in the bulk states are the VTBICs of $h^{(2)}$ and $h^{(3)}$ subsystems (magenta and cyan solid curves). Black dashed curve denotes the trivial interface states of $h^{(3)}$ subsystem. Right panel: Distributions of eigenstates marked by the green and yellow stars in the left panel. The green lines indicate the interfaces. The parameters are chosen as $m = 0.5$, $m_5 = -1.3$, $t_a = -0.4$, $t_b = -1.6$, $t_1 = -1.2$, and $t_2 = -1.7$.

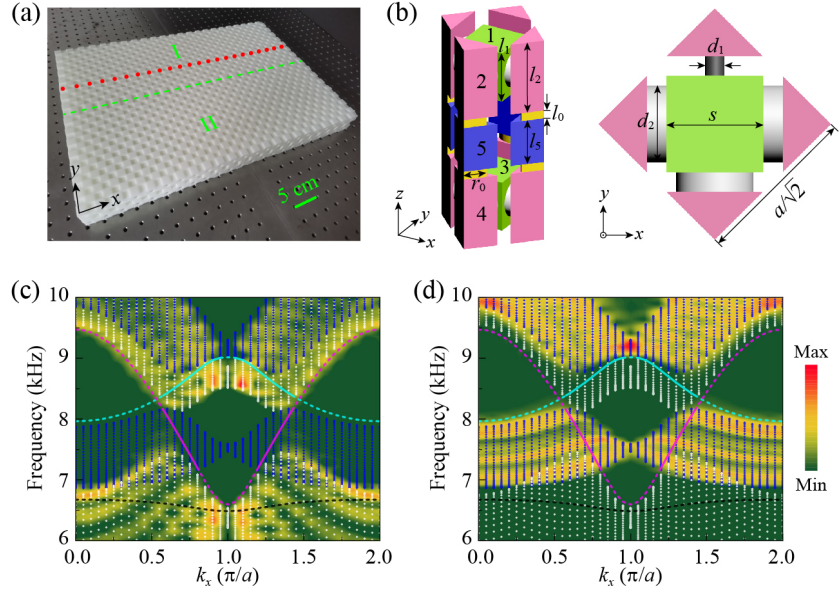

Fig. 2. Acoustic realizations of VTBICs. (a) Photograph of the PC sample. (b) Side and top views of the unit cell of phase I. (c) Projected dispersions of the PC in (a), which is composed of two opposite topological phases, i.e., phase I and phase II. Again, the magenta (cyan) solid curves represent the VTBICs from the $h^{(2)}$ ($h^{(3)}$) subsystem lying in the extended bulk states of the $h^{(3)}$ ($h^{(2)}$) subsystem denoted by blue (white) dots. The color map denotes the measured projected dispersions of bulk states of the $h^{(2)}$ subsystem, which captures precisely the simulated results (white dots). (d) The same to (c), but the color map denotes the measured projected dispersions of bulk states of the $h^{(3)}$ subsystem.

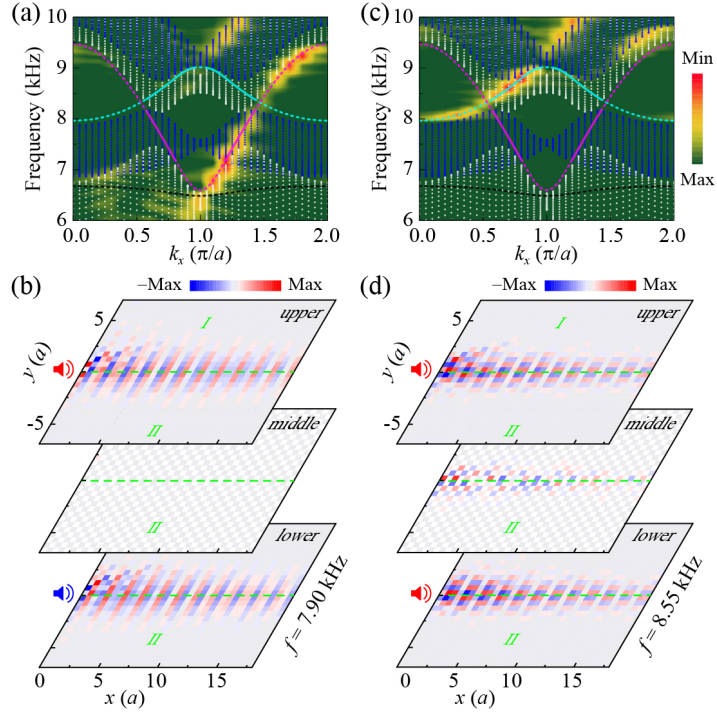

Fig. 3. Observations of VTBICs. (a) and (b) Measured projected dispersions and pressure field distributions of VTBICs from the $h^{(2)}$ subsystem. The interface of the two phases is indicated by green lines. The red and blue microphones present two point sources with $0$ and $\pi$ phases, respectively. (c) and (d) The same to (a) and (b), but for the VTBICs from the $h^{(3)}$ subsystem.

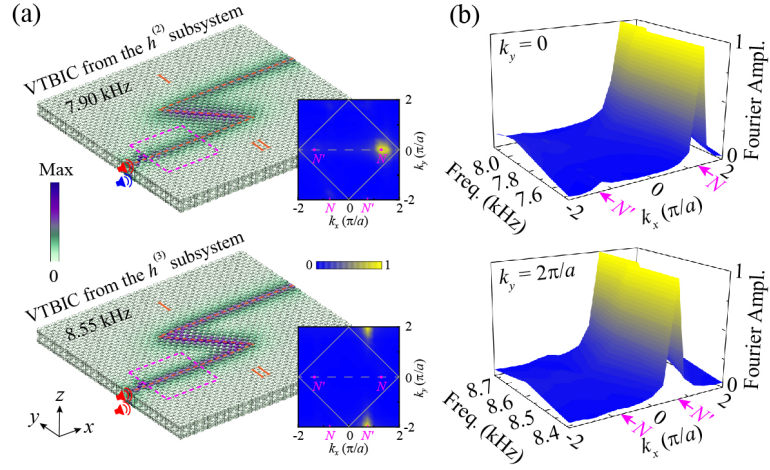

Fig. 4. Robustness of the VTBICs. (a) Simulated pressure field distributions for the VTBICs from the $h^{(2)}$ (top panel) and $h^{(3)}$ (bottom panel) subsystems propagating along the Z-shaped channel. Insets: Measured spatial Fourier spectra for the VTBICs, which are performed within the domains marked by magenta rectangles. (b) Top panel: Magnitude of experimental Fourier spectra of $k_y = 0$ in the frequency range that only the VTBICs from the $h^{(2)}$ subsystem can exist. Bottom panel: The same to the top panel, but for the spectra of $k_y = 2\pi/a$ for VTBICs from the $h^{(3)}$ subsystem.